\documentclass[prl,twocolumn,showpacs,floatfix,amssymb,superscriptaddress]{revtex4}

\usepackage{amsmath}
\usepackage{amsfonts}
\usepackage{graphicx}

 \DeclareMathOperator{\Ren}{Re}
\DeclareMathOperator{\Imn}{Im} 

\begin{document}

\title{Coulomb Blockade and Super Universality of the $\theta$ Angle}
\author{I.S.\,Burmistrov}
\affiliation{L.D. Landau Institute for Theoretical
Physics RAS, Kosygina street 2, 119334 Moscow, Russia}
\affiliation{Department of Theoretical Physics, Moscow
Institute of Physics and Technology, 141700 Moscow, Russia}
\author{A.M.M.\,Pruisken}
\affiliation{Institute for Theoretical Physics, University
of Amsterdam, Valckenierstraat 65, 1018XE Amsterdam, The
Netherlands }
\begin{abstract}

Based on the Ambegaokar-Eckern-Sch\"{o}n approach to the Coulomb
blockade we develop a complete quantum theory of the single
electron transistor. We identify a previously unrecognized
physical observable ($q^\prime$) in the problem that, unlike the
usual average charge ($Q$) on the island, is robustly quantized
for any {\em finite} value of the tunneling conductance as the
temperature goes to absolute zero. This novel quantity is
fundamentally related to the non-symmetrized noise of the system.
We present a unifying scaling diagram in the $q^\prime$ -
$g^\prime$ plane where $g^\prime$ denotes the conductance of the
system. The results display all the super universal topological
features of the $\theta$ angle concept that previously arose in
the theory of the quantum Hall effect.

\end{abstract}

 \maketitle


 The Ambegaokar-Eckern-Sch\"{o}n
(AES) model~\cite{AES} is the simplest approach to the Coulomb
blockade problem ~\cite{SZ,SET} that has attracted a considerable
amount of interest over the years. The standard experimental
set-up is the single electron transistor (SET) which is a
mesoscopic metallic island coupled to a gate and connected to two
metallic reservoirs by means of tunnelling contacts with a total
conductance $g$. Even though the physical conditions of the AES
model are limited and well known,~\cite{Falci,BEAH,Glazman} the
theory nevertheless displays richly complex and fundamentally new
behavior, much of which has not been understood to date. To
explain the observed tunnelling phenomena with varying temperature
$T$ and gate voltage $V_g$ one usually considers an isolated
island or a {\em single electron box} obtained by putting the
tunnelling conductance $g$ equal to zero. The AES model then leads
to the standard semiclassical or electrostatic picture of the
Coulomb blockade which says that at $T=0$ the average charge ($Q$)
on the island is {\em robustly} quantized except for very special
values of the gate voltage $V_g^{(k)}=e(k+1/2)/C_g$ with integer
$k$ and $C_g$ denoting the gate capacitance. At these very special
values a {\em quantum phase transition} occurs where the average
charge $Q$ on the island changes from $Q=k$ to $Q=(k+1)$ in units
of $e$.

The experiments on the SET always involve {\em finite} values of
the tunnelling conductance $g$, however, and this dramatically
complicates the semiclassical picture of the Coulomb blockade.
Despite ample theoretical work on both the strong and weak
coupling side of the problem, the matter still lacks basic
physical clarity since the averaged charge $Q$ is known to be {\em
un}-quantized for any finite value of $g$, no matter how
small.~\cite{Matveev} This raises fundamental questions about the
exact meaning of the experiments and the physical quantities in
which the Coulomb blockade is usually expressed.

In this Letter we present a complete quantum theory of the SET
that is motivated by the formal analogies that exist between the
AES theory on the one hand, and the theory of the quantum Hall
effect ~\cite{PruiskenBurmistrov2} on the other. Each of these
theories describe an interesting experimental realization of the
topological issue of a $\theta$ vacuum that originally arose in
QCD.~\cite{Rajaraman} In each case one deals with different
physical phenomena and therefore different quantities of physical
interest. What has remarkably emerged over the years is that the
basic scaling behavior is always the same, independent of the
specific application of the $\theta$ angle that one is interested
in.~\cite{PruiskenBurmistrov2} Within the grassmannian
$U(m+n)/U(m) \times U(n)$ non-linear $\sigma$ model, for example,
one finds that quantum Hall physics, in fact, a {\em super
universal} topological feature of the theory for all values of $m$
and $n$. It is therefore of interest to know whether super
universality is retained in the AES theory where physical concepts
such as the ``Hall conductance" and ``$\theta$ renormalization"
have not been recognized.

In direct analogy with the theory of the quantum Hall effect we
develop, in the first part of this Letter, a quantum theory of
{\em observable parameters} $g^\prime$, $E_c^\prime$ and
$q^\prime$ for the AES model obtained by studying the {\em
response} of the system to changes in the boundary conditions.
Here, $g^\prime$ is identified as the SET conductance,
$E_c^\prime$ is the charging energy whereas the $q^\prime$ is a
novel physical quantity that is fundamentally related to the
current noise in the SET. The $q^\prime$ is in all respects the
same as the Hall conductance in the quantum Hall effect and,
unlike the averaged charge $Q$ on the island, it is robustly
quantized in the limit $T \rightarrow 0$. The crux of this Letter
is the unifying scaling diagram of Fig. 1 indicating that
$g^\prime$ and $q^\prime$ are the appropriate renormalization
group parameters of the AES theory. These scaling results which
are the main objective of the remaining part of this Letter,
provide the complete conceptual framework in which the various
disconnected pieces of existing computational knowledge of the AES
theory can in general be understood.

\begin{figure}[t]
\includegraphics[height=65mm]{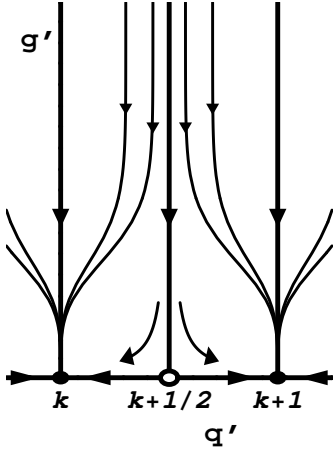}
\vspace{-.5cm} \caption{Unified scaling diagram of the Coulomb
blockade in terms of the SET conductance $g^\prime$ and the quasi
particle charge $q^\prime$. The arrows indicate the scaling toward
$T = 0$ (see text).} \label{RGFIG}\vspace{-.5cm}
\end{figure}

\noindent{\bf
\textsc{AES model}.} The action involves a single abelian phase
$\phi(\tau)$ describing the potential fluctuations on the island
$V(\tau)=i\dot{\phi}(\tau)$ with $\tau$ denoting the imaginary
time.~\cite{AES} The theory is defined by
\begin{equation}\label{Zstart}
 Z = \int \mathcal{D}[\phi] e^{-\mathcal{S} [\phi]}, \quad \mathcal{S} [\phi] =
 \mathcal{S}_d +\mathcal{S}_t + \mathcal{S}_c.
\end{equation}
The action $\mathcal{S}_d$ describes the tunneling between the
island and the leads
\begin{equation}\label{SdStart}
 \mathcal{S}_d[\phi] = \frac{g}{4}\int_{0}^{\beta} d\tau_1 d\tau_2
 \alpha(\tau_{12})e^{-i[ \phi(\tau_1)-\phi(\tau_2)]}
\end{equation}
where $\beta=1/T$, $\tau_{12} = \tau_1-\tau_2$ and the kernel
$\alpha(\tau)$ is usually expressed as
$\alpha(\tau)=(T/\pi)\sum_{n}|\omega_n|e^{-i\omega_n\tau}$ with
$\omega_n=2\pi T n$. The part $\mathcal{S}_t$ describes the
coupling between the island and the gate and $\mathcal{S}_c$ is
the effect of the Coulomb interaction between the electrons
\begin{equation}\label{StcStart}
 \mathcal{S}_t[\phi] = - 2\pi i q \mathcal{C} [\phi] , \quad
 \mathcal{S}_c[\phi] = \frac{1}{4 E_c} \int_{0}^{\beta} d \tau\, \dot{\phi}^2 .
\end{equation}
Here, $q=C_g V_g/e$ is the external charge and $\mathcal{C} [\phi]
= 1/(2\pi)\int_{0}^{\beta}d\tau \dot{\phi}$ is the winding number
or {\em topological charge} of the $\phi$ field. For the system in
equilibrium the winding number is strictly an integer~\cite{BEAH}
which means that Eq.~\eqref{StcStart} is only sensitive to the
{\em fractional} part $-\frac{1}{2} < q < \frac{1}{2}$ of the
external charge $q$. The main effect of $\mathcal{S}_c$ in
Eq.~\eqref{StcStart} is to provide a cut-off for large
frequencies. Eq.~\eqref{SdStart} has classical finite action
solutions $\phi_0 (\tau)$ with a non-zero winding number that are
completely analogous to Yang-Mills instantons. The general
expression for winding number $W$ is given by
~\cite{Instantons1,Instantons2}
\begin{equation}
 e^{i\phi_{0}(\tau)} = e^{-i 2\pi T\tau} \sum \limits_{a=1}^{|W|}\frac{e^{i 2\pi
 T\tau}-z_a}{e^{-i 2\pi T\tau}-{z}_a^*}.\label{InstSolG}
\end{equation}
For instantons ($W>0$) the complex parameters $z_a$ are all inside
the unit circle and for anti-instantons ($W<0$) they are outside.
Considering $W=\pm 1$ which is of interest to us, one identifies
$\textrm{arg}\, z /2\pi T$ as the {\em position} (in time) of the
single instanton whereas $\lambda=(1-|z|^2)\beta$ is the {\em
scale size} or the duration of the potential pulse
$i\dot{\phi}_{0}(\tau)$. The semiclassical instanton expression
for the thermodynamic potential $\Omega=-T \ln Z$ can be written
in the standard form ~\cite{Instantons3}
\begin{equation}
 {\Omega_\textrm{inst}} = - g D \int_0^{\beta}
 \frac{d\lambda}{\lambda^2} e^{-\frac{g(\lambda)}{2} - \mathcal{O} (
 1/\lambda E_c(\lambda ))} \cos 2\pi q .\label{Oinst}
\end{equation}
Here, $D = 2 e^{-\gamma_E-1}$ with $\gamma_E\approx 0.577$
denoting the Euler constant. Introducing a frequency scale $\nu_0
= g E_c/(\pi^2 D)$ then $g(\lambda)$ and $E_c(\lambda)$ are given
by ~\cite{perturb}
\begin{equation}
 g(\lambda) = g - 2 \ln \lambda \nu_0,\,\,\,  E_c(\lambda) = E_c
 \Bigl [1 - \frac{2}{g} \ln \lambda \nu_0 \Bigr ] .\label{Pert}
\end{equation}
The logarithmic corrections are the same as those computed in
ordinary perturbation theory in $1/g$. Based on Eq.~\eqref{Pert}
alone one expects that the SET always scales from a good {\em
conductor} at high $T$ or short times $\lambda\nu_0 \ll 1$ to an
{\em insulator} at low $T$ or long times $\lambda\nu_0 \gg 1$.

\noindent{\bf
\textsc{Kubo formulae.}} To develop a theory of {\em observable}
parameters of the SET~\cite{PruiskenBurmistrov2} we employ the
background field $\tilde{\phi} (\tau)=\omega_n \tau$ that
satisfies the classical equation of motion of Eq.~\eqref{Zstart}.
Write
\begin{equation}
 e^{-\mathcal{S}^{\prime} [\tilde{\phi}]} =
 Z^{-1} \int \mathcal{D}[\phi] e^{-\mathcal{S} [\tilde{\phi} + \phi]}
 \label{Zstart1}
\end{equation}
then a detailed knowledge of $\mathcal{S}^{\prime} [\tilde{\phi}]$
generally provides complete information on the low energy dynamics
of the system. The effective action
$\mathcal{S}^{\prime}[\tilde{\phi}]$ is properly defined in terms
of a series expansion in powers of $\omega_n$. Retaining only the
lowest order terms in the series we can write
\begin{equation}\label{Seff}
 \mathcal{S}^\prime [\tilde{\phi} ]=\beta\left ( \frac{g^\prime}{4\pi}
 |\omega_n| - i q^\prime \omega_n+\frac{\omega_n^2}{4
 E_c^\prime}\right ).
\end{equation}
The quantities of physical interest are $g^\prime$, $q^\prime$
with $|q^\prime| < 1/2$ and $1/E_c^\prime$ that are formally given
in terms of Kubo-like expressions~\cite{Det}
\begin{gather}
 g^\prime = 4\pi \Imn {\langle K(\eta)\rangle}\Bigr |_{\eta \rightarrow 0},\,
 q^\prime = q + \frac{i\langle\dot{\phi}\rangle}{2E_c}+
 \Ren {\langle K(\eta)\rangle} \Bigr |_{\eta \rightarrow 0}
 ,\notag\\
 \frac{1}{E_c^\prime} =\frac{1}{E_c} \left( 1 +
 \int_0^\beta d\tau e^{i\eta\tau}{\left \langle \dot{\phi}(\tau)
 K(\eta)\right \rangle} \right) \Bigr |_{\eta \rightarrow 0}  \label{Back}
\end{gather}
where the expectation is with respect to the theory of
Eq.~\eqref{Zstart}. Here, $K(\eta)$ is obtained from the
expression
\begin{equation}
 K(i\omega_n)=\frac{g}{4\beta}\int_0^\beta {d\tau_1
 d\tau_2} \left[ \frac{e^{i\omega_n\tau_{12}}-1}{i\omega_n} \right] \alpha(\tau_{12})
 e^{i[{\phi(\tau_2)-\phi(\tau_1)}]} \label{Kdef}
\end{equation}
followed by the analytic continuation $i\omega_n\to \eta+i0^+$
which is standard. As we shall point out in what follows, the main
advantage of the background field formalism of Eqs
\eqref{Zstart1}-\eqref{Kdef} is that it unequivocally determines
the renormalization of the AES model while retaining the close
contact with the physics of the SET. To see this we notice first
that by expanding the effective action of Eq. \eqref{Seff} in
powers of $\omega_n$ we essentially treat the discrete variable
$\omega_n$ as a continuous one. This means that the quantities
$g^\prime$, $q^\prime$ with $|q^\prime| < 1/2$ and $1/E_c^\prime$
in Eqs~\eqref{Back} -\eqref{Kdef} are, by construction, a measure
for the response of the system to infinitesimal changes in the
boundary conditions. This observation immediately leads to a
general criterion for the strong coupling {\em Coulomb blockade
phase} of the SET that the perturbative results of Eq.
\eqref{Pert} could not give. More specifically, the general
statement which says that the SET scales toward an {\em insulator}
as $T \rightarrow 0$ implies that the response quantity
$g^\prime$, the fractional part of $q^\prime$ as well as the
dimensionless quantity $1/\beta E^\prime_c$ all render equal to
zero except for corrections that are exponentially small in
$\beta$. Since the expressions of Eqs~\eqref{Back} -\eqref{Kdef}
are all invariant under the shift $q \to q +k$ and $q^\prime \to
q^\prime +k$ for integer $k$, we conclude that the AES theory on
the strong coupling side generally displays the {\em Coulomb
blockade} with the novel quantity $q^\prime$, unlike the averaged
charge $Q$ on the island, now identified as the {\em robustly
quantized} quasi particle charge of the SET. This quantization
phenomenon which is depicted in Fig.~\ref{RGFIG} by the infrared
stable fixed points located at integer values $q^\prime =k$, is
fundamentally different from semiclassical picture of the Coulomb
blockade since it elucidates the discrete nature of the electronic
charge which is independent of tunneling.

Before embarking on the details of scaling it is important to
emphasize that Eqs~\eqref{Seff} -\eqref{Kdef} are precisely the
same quantities that one normally would obtain in ordinary linear
response theory. For example, $g^\prime$ is exactly same as the
Kubo formula~\cite{Cur} relating a small potential difference $V$
between the leads to the current $I$ across the island according
to $I = e^2 G V/h$ where $G = g_l g_r g^\prime/(g_l+g_r)^2$. Here,
$h$ is Planck's constant and $g_{l,r}$ are the bare tunneling
conductances across the different leads. To understand the new
quantity $q^\prime$ we notice that the first two terms in Eq.
\eqref{Back} are equal to the average charge $Q$ on the island,
$Q=q-(2E_c)^{-1}\partial \Omega/\partial q$. The last piece in
$q^\prime$ is related to the current noise~\cite{Det} and a more
transparent expression is obtained by writing
\begin{equation}
q^\prime = Q + \frac{(g_l+g_r)^2}{2g_l g_r}\left . i
\frac{\partial}{\partial V} \int_0^\infty dt \langle  [I(t),
I(0)]\rangle \right |_{V=0}.\label{CorI}
\end{equation}
It can be shown that the last term in Eq.~\eqref{CorI} is the
result of an inductive coupling between tunneling current of the
SET and the external current in the circuit. Similarly, it can be
shown that $E_c^\prime$ describes the frequency dependence of the
tunneling current ($I_\omega$) and the potential difference
($V_\omega$) according to $\partial
(I_\omega/V_\omega)/\partial\omega = ie^2/(2 E_c^\prime)$.

\noindent{\bf
\textsc{Weak coupling phase.}} By evaluating Eqs~\eqref{Back}
-\eqref{Kdef} in a series expansion in powers of $1/g$ one obtains
the same lowest order results as in Eq.~\eqref{Pert} but with
$q^\prime = 0$. The quantity $q^\prime$ is generally unaffected by
the quantum fluctuations and to establish the renormalization of
$q^\prime$ it is necessary to include instantons. Following the
detailed methodology of Ref.~[\onlinecite{PruiskenBurmistrov2}] we
express the observable theory in terms of renormalization group
$\beta = \beta(g^\prime,q^\prime)$ and $\gamma = \gamma
(g^\prime,q^\prime)$ functions which are universal and given by
~\cite{Det}
\begin{eqnarray}
\beta_g &=& \frac{d g^\prime}{d \ln \lambda} = -2 -
\frac{4}{g^\prime} - D g^{\prime 2} e^{-g^\prime/2} \cos 2\pi
q^\prime \hspace{0.5cm}   \label{NPRG1}
\\
\beta_q &=& \frac{d q^\prime}{d\ln \lambda}  = ~~~~~~~~~~ -
\frac{D}{4\pi} g^{\prime 2} e^{-g^\prime/2} \sin 2\pi q^\prime
\label{NPRG2} \\
\gamma &=& \frac{d \ln E_c^\prime}{d\ln \lambda} =~-
\frac{2}{g^\prime} + \frac{D}{2}~ g^{\prime 2} e^{-g^\prime/2}
\cos 2\pi q^\prime .\label{NPRG3}
\end{eqnarray}
Here, $D$ is the same as in Eq.~\eqref{Oinst} and we have included
in Eq. \eqref{NPRG1} the perturbative contribution of order
$1/g^\prime$.~\cite{HZ} The results indicate that instantons are
the fundamental objects of the theory that facilitate the {\em
cross-over} between the metallic phase with $g^\prime \gg 1$ at
high $T$ and the Coulomb blockade phase with $g^\prime \lesssim 1$
that generally appears at a much lower $T$ only.

\noindent{\bf
\textsc{Strong coupling phase.}} We next evaluate Eqs~\eqref{Back}
-\eqref{Kdef} in terms of a strong coupling expansion about the
theory with $g=0$.~\cite{SZ,DAR} Remarkably, this expansion is in
many respects the same as the one recently reported for the two
dimensional $CP^{N-1}$ model with large values of $N$.~\cite{CPN}
The results for small values of $g^\prime$ and $u^\prime= q^\prime
- k - 1/2$ can be written as follows ~\cite{Det}
\begin{equation}
\beta_g = -\frac{g^{\prime 2}}{\pi^2},\,\quad \beta_q =
u^\prime\left (1-\frac{g^\prime}{\pi^2}\right ),\,\quad \gamma
=O(g^{\prime 2}) \label{SCb}
\end{equation}
indicating that $u^\prime = g^\prime=0$ is the {\em critical}
fixed point of the AES theory with $g^\prime$ a marginally
irrelevant scaling variable. Eq. \eqref{SCb}, together with the
weak coupling results of Eqs \eqref{NPRG1} - \eqref{NPRG3}, are
the main justification of the unifying scaling theory illustrated
in Fig.~\ref{RGFIG}. To make contact with the existing strong
coupling analysis of $g^\prime$ ~\cite{Matveev,Falci,Schon} we
employ the basic principles of the renormalization group and
obtain the general scaling results $g^\prime = g^\prime (X,Y)$ and
$q^\prime = q^\prime (X,Y)$ where~\cite{Det}
\begin{equation}
X = \tilde{E}_c \tilde{u} /(T \tilde{g}) ,~Y = (T/\tilde{E}_c)
e^{-1/\tilde{g}}
\end{equation}
with $\tilde{u}$, $\tilde{g}$ and $\tilde{E}_c$ denoting the
renormalization group starting point (which, by the way, is
slightly different from the bare theory $u$, $g$ and $E_c$). An
explicit computation gives $g^\prime(0,Y) = |\ln Y|^{-1}$
indicating that the maximum of $g^\prime$ decreases with $T$ like
$|\ln T|^{-1}$. Similarly we find $q^\prime(X,Y) = k+1/2-X|\ln
Y|^{-1}$ indicating that the width $\Delta V_g $ of the transition
with varying $V_g \propto q$ vanishes with $T$ according to
$\Delta V_g \propto T |\ln T|$.~\cite{Schon}

\noindent{\bf
\textsc{Critical correlations.}} Of general interest are the
critical correlations of the AES theory with $u, g \approx 0$.
These are most elegantly described by the fermionic effective
action \cite{CPN}
\begin{eqnarray}
 \mathcal{S} = \int \bar{\psi} ( \partial_\tau  + b
 \sigma_z ) \psi ~ + \frac{g}{4} \int_{1,2}
 S_-(\tau_1) \alpha (\tau_{12}) S_+(\tau_2) .~~~~\label{SeffCrit}
\end{eqnarray}
Here, $b=E_c u \approx 0$, $\sigma_{x,y,z}$ are the Pauli matrices
and $\bar{\psi}$ and $\psi$ denote two component fermion fields.
The operators $S_{\pm} (\tau) =\frac{i}{2} \bar{\psi} (\tau)
(\sigma_x \pm i \sigma_y ) \psi (\tau)$ in Eq. \eqref{SeffCrit}
are identified with the AES operators $e^{\pm i\phi (\tau)}$ in
Eqs \eqref{Zstart} - \eqref{StcStart} that create (annihilate) a
unit charge on (from) the island at time $\tau$. In the absence of
tunneling ($g=0$) one has $\langle \bar{\psi} \sigma_z \psi
\rangle = b/|b|$ indicating that the transition at $b=0$ is a
first order one. Furthermore,
\begin{eqnarray}
 \langle e^{i\{\phi (0) - \phi (\tau)\} }  \rangle &=&
 \langle S_{+} (0) S_{-} (\tau)  \rangle =
 \vartheta(b \tau/|b|) e^{-2 |b| \tau} ~~~~~\label{charge-cre}
\end{eqnarray}
where $\vartheta$ is the Heaviside step function. These results
are precisely in accordance with the semiclassical picture of the
SET where $2 |b|$ denotes the {\em continuously vanishing} energy
gap between the states $q^\prime =Q = k $ and $k+1$ of the island
as one approaches the critical point. In the presence of tunneling
$g \ne 0$, however, the correlations get more complicated and the
main technical problem is to find the modified operators $S_\pm$
that change the quasi particle charge $q^\prime$ of the SET rather
than the averaged charge $Q$ of the island.~\cite{Det}

In summary, based on the new concept of $\theta$ or $q^\prime$
renormalization we assign a universal significance to the Coulomb
blockade in the SET that previously did not exist beyond the
semiclassical picture. We have shown that the AES model is, in
fact, an extremely interesting and exactly solvable example of a
$\theta$ vacuum that displays all the {\em super universal}
topological features that have arisen before in the context of the
quantum Hall liquids~\cite{PruiskenBurmistrov2} as well as quantum
spin liquids.~\cite{Spin} These include not only the existence of
{\em gapless} or critical excitations at $q^\prime=k+1/2$ (or
$\theta = \pi$) but also the {\em robust} topological quantum
numbers that explain quantization of the electronic charge in the
SET at finite values of $g$. Unlike the conventional theories of
the $\theta$ angle, however, the strong coupling behavior of the
AES model can be studied analytically and our results for the
novel quantity $q^\prime$ should in general be taken as an
experimental challenge. Following Eq.~\eqref{CorI} it involves the
anti-symmetric part of the current noise which can be
experimentally observed by a coupling of the SET to an $LC$
component,~\cite{LesLoos} for example, see also Ref.
[\onlinecite{Blanter}].

Notice that the critical behavior of the SET is likely to change
when the number of channels in the tunneling contacts are taken to
be finite rather than infinite.~\cite{Matveev} Under these
circumstances one expects a {\em second order} transition at
$q^\prime = k+1/2$~\cite{Matveev} with a finite value of the SET
conductance $g^\prime$ which closely resembles the more
complicated physics of the quantum Hall
effect.~\cite{PruiskenBurmistrov2} Finally, the AES theory is
known to map onto the ``circular brane'' model~\cite{LZ} such that
the findings of this Letter apply to the latter theory as well. It
should be mentioned that physical objectives similar to ours have
recently been pursued in Ref.[\onlinecite{Bulgadaev}] using
otherwise heuristic arguments. The reported ideas and conjectures,
however, are in many ways in conflict with the present theory.

The authors are indebted to A. Abanov for bringing the AES model
to their attention as well as A. Lebedev and Yu. Makhlin for
helpful discussions. The research was funded in part by
\textit{CRDF}, the Dutch Science Foundations \textit{NWO} and
\textit{FOM} as well as \textit{RFBR}, \textit{RSSF},\textit{MES},
\textit{FASI} and \textit{RAS} of Russia.


\end{document}